\documentclass{aip-cp}

\usepackage[numbers]{natbib}
\usepackage{rotating}
\usepackage{graphicx}
\usepackage{bm}

\begin{document}

\title{Fully coupled-channel study of $K^-pp$ resonance in a chiral SU(3)-based $K^{bar}N$ potential}

\author[aff1,aff2]{Akinobu Dot\'e\corref{cor1}}
\author[aff3]{Takashi Inoue}
\author[aff4]{Takayuki Myo}

\affil[aff1]{KEK Theory Center, Institute of Particle and Nuclear Studies (IPNS), High Energy Accelerator Research Organization (KEK), 1-1 Oho, Tsukuba, Ibaraki, 305-0801, Japan}
\affil[aff2]{J-PARC Branch, KEK Theory Center, IPNS, KEK, 203-1, Shirakata, Tokai, Ibaraki, 319-1106, Japan}
\affil[aff3]{Nihon University, College of Bioresource Sciences, Fujisawa 252-0880, Japan}
\affil[aff4]{General Education, Faculty of Engineering, Osaka Institute of Technology, Osaka 535-8585, Japan}
\corresp[cor1]{Corresponding author: dote@post.kek.jp}

\maketitle

\begin{abstract}
Nuclear system with antikaons, so-called kaonic nuclei, has been a longstanding issue in strange nuclear physics and hadron physics, because they might have exotic nature; In particular, they could be a doorway to the dense matter due to the strong attraction between antikaon and nucleon. Among kaonic nuclei, the three-body system composed of two protons and a single $K^-$ meson, $K^-pp$, is the most essential. In this article\footnote{It is remarked that this article is mainly based on our latest study \cite{Full-ccCSM_Chiral:Dote}, since organizers of the conference asked us to give a talk on it.}, we will report on the recent situation of $K^-pp$ studies in both theoretical and experimental sides. Afterwards, we will explain our latest study of the $K^-pp$ with a fully coupled-channel complex scaling method (Full ccCSM) using a chiral SU(3)-based $\bar{K}N$(-$\pi Y$) potential. In Full ccCSM, the $K^-pp$ is  completely treated as a resonant state of a $\bar{K}NN$-$\pi\Sigma N$-$\pi\Lambda N$ coupled-channel system. The energy dependence involved in the chiral potential is handled with a self-consistent calculation in which two extreme ansatzes are examined: field picture and particle picture. With our chiral SU(3)-based potential constrained by the precise data of kaonic hydrogen atom (SIDDHARTA experiment), the $K^-pp$ resonance is obtained as a shallowly bound state measured from the $\bar{K}NN$ threshold with a narrow width, if the field picture is employed in the calculation: the binding energy is $14-28$ MeV and half value of the mesonic decay width are $8-15$ MeV. On the other hand, if the particle picture is employed, it is found that the binding energy could be as large as about 50 MeV, even though such a chiral potential is used. Based on these results of Full ccCSM calculation, we have discussed on the possibility for kaonic nuclei to form a dense matter and on the latest experimental result reported by J-PARC E15 collaboration.
\end{abstract}

\section{INTRODUCTION}

As a nuclear many-body system with strangeness, kaonic nuclei have been one of the important topics in strange nuclear physics and hadron physics. Kaonic nuclei are nuclear system with antikaons (a pseudoscalar meson $\bar{K}=(K^-, \bar{K}^0)$). Kaonic nuclei are an unique system since real mesons are contained  as a constituent of the system, not a virtual meson mediating the interaction. Such kaonic nuclei are expected to have several interesting properties that we have never seen in ordinary nuclei, due to the strong attraction between an antikaon and a nucleon. Nowadays, an excited hyperon $\Lambda(1405)$ is recognized as a quasi-bound state of an antikaon $\bar{K}$ and a nucleon $N$ (or a meson-baryon molecule state), rather than a genuine three-quark state \cite{ChU:Review}. Accordingly, the $\bar{K}N$ attraction, especially in the isoscalar channel, is considered to be so strong to form a quasi-bound state which corresponds to the $\Lambda(1405)$. It is noted that a recent study with a lattice QCD calculation has supported such a molecular nature of the $\Lambda(1405)$: They have found that the $\Lambda(1405)$ is dominated by the $\bar{K}N$ component when the pion mass is so small as that of experimental value \cite{L1405-LatticeQCD}. Such a strong $\bar{K}N$ attraction is considered to bring exotic natures to kaonic nuclei, as suggested by earlier studies using an phenomenological potential \cite{AY_2002}. Owing to the strong attraction from antikaons, nucleons are attracted close to the antikaons against the $NN$ repulsive core. As a result, a dense state with several times higher than the normal nuclear density might be formed in kaonic nuclei \cite{AMDK}. Therefore, kaonic nuclei could be a doorway to the dense matter. In hadron physics, a partial restoration of chiral symmetry has been an important theme for a long time, which might occur in hot and/or dense matter \cite{ChSymRes:Hatsuda, ChSymRes:Weise}. Kaonic nuclei are expected to give some hints to this longstanding issue. 

Thus, kaonic nuclei are expected to be an exotic system. In order to reveal the properties of kaonic nuclei, many researchers have focused on the $K^-pp$, which is a three-body system composed of a single $K^-$ meson and two protons. Certainly, general kaonic nuclei might possess interesting properties. However, it is difficult to investigate such a strongly-interacting quantum many-body system. Therefore, before proceeding to general systems, the simple three-body system, $K^-pp$, has been investigated eagerly from both theoretical and experimental sides. As mentioned in the previous paragraph, the $\Lambda(1405)$ is a quasi-bound state of two-body $\bar{K}N$ system. In our viewpoint, the $\Lambda(1405)$ is a building block of kaonic nuclei, and the $K^-pp$ is a prototype of kaonic nuclei, which means the most essential kaonic nucleus. The $K^-pp$ could be a bridge from the excited hyperon $\Lambda(1405)$ to general systems of kaonic nuclei. 

The current situation of the $K^-pp$ study is as follows. There have been many experiments to search for the $K^-pp$ quasi-bound states. However, the results are different in each experiment. In an early experiment by FINUDA collaboration \cite{Kpp:exp_FINUDA}, a bump structure was reported in the observed $\Lambda p$ invariant-mass distribution at far below the $\bar{K}NN$ threshold. In addition, DISTO \cite{Kpp:exp_DISTO} and J-PARC E27 \cite{Kpp-ex:JPARC-E27} collaborations reported a signal around $\pi\Sigma N$ threshold, which exists 103 MeV below the $\bar{K}NN$ threshold. In a naive consideration, these three experiments indicate a deeply bound state of $K^-pp$ with about 100 MeV binding energy. On the other hand, J-PARC E15 collaboration, in their first run \cite{Kpp-ex:JPARC-E15}, reported a signal in a shallowly bound region, which indicates the $K^-pp$ binding energy is so small as about 15 MeV. In addition, no signal was found in a experiment by LEPS/SPring8 group \cite{Kpp-ex:LEPS}. However, there are several controversies on these experiments \cite{Kpp_Criticism:Magas, Kpp-ex:HADES-PWA, EXA14:Gal}. Interestingly, a theoretical study suggested that the signals observed in DISTO and J-PARC E27 could be a weakly bound state of $\pi\Sigma N$ coupled with $\pi\Lambda N$ (so-called pion-assisted dibaryon), not a $\bar{K}NN$ bound state, since it is close to the $\pi\Sigma N$ threshold \cite{EXA14:Gal}. It should be noted that the J-PARC E15 collaboration has recently reported the result of the second-run experiment of $^3$He(inflight $K^-$, $\Lambda p$)$n_{missing}$ with high statistics \cite{Kpp-ex:JPARC-E15-2nd_fin}. The shape of the observed spectrum is quantitatively consistent with a theoretical prediction \cite{JPARC-E15:Sekihara-Oset-Ramos}, which shows a peak below and above the $\bar{K}NN$ threshold (two-peak structure). According to the analysis by the J-PARC E15 collaboration, it is reported that the binding energy of $K^-pp$ is $47 \pm 3 (stat.)\; ^{+3}_{-6} (sys.)$ MeV and the decay width is $115 \pm 7 (stat.)\; ^{+10}_{-9} (sys.)$ MeV. 

As for the theoretical side, there are many studies of the $K^-pp$ using various methods, because the $K^-pp$ is just a three-body system. In most of theoretical studies, the $K^-pp$ has been investigated with a variational method or Faddeev-Alt-Grassberger-Sandhas (Faddeev-AGS) approach using a phenomenological $\bar{K}N$ potential or a chiral theory-based $\bar{K}N$ potential. In those calculations, the $K^-pp$ is considered as a $\bar{K}NN$-$\pi\Sigma N$-$\pi\Lambda N$ coupled-channel system with quantum numbers of spin-parity $J^\pi=0^-$ and isospin $T=1/2$. Hereafter, we denote such a $K^-pp$ state as $``K^-pp"$ ($K^-pp$ with double quotation marks). It is noted that with these quantum numbers, the strong attraction of the $\bar{K}N$ potential in the isoscalar channel can be mostly gained in the three-body system of an antikaon and two nucleons. As summarized in Table XVII of Ref. \cite{StrangenessSummary_2016}, the calculated values of binding energy are scattered, depending on the employed potential model. In both calculations of variational method and Faddeev-AGS approach, the binding energy of $``K^-pp"$ is obtained to be so large as 50-90 MeV with phenomenological potentials, which are energy-independent \cite{Kpp:AY, Faddeev:Shevchenko}. On the other hand, it is obtained to be rather small as $~20$ MeV with chiral SU(3) potentials, which are energy-dependent \cite{Kpp:DHW, Kpp:BGL, Kpp:IKS}. Thus, the consensus in the theoretical side is that the $``K^-pp"$ should be bound with the binding energy of less than 100 MeV, but that its value strongly depends on employed $\bar{K}N$ potential models.

Under such a current situation, we have carried out a novel calculation with a fully coupled-channel complex scaling method \cite{Full-ccCSM_Chiral:Dote, Full-ccCSM_AY:Dote}, in which we care all the important ingredients for the theoretical study of the $K^-pp$. This article is organized as follows: In the next section, the methodology of our study will be explained in detail. Then, the results of the $``K^-pp"$ obtained with our method are shown, and there are several discussions based on our results. In the last section, the summary and future prospects are described.

\section{METHODOLOGY}

In this section, we explain all ingredients used in the present study of the ``$K^-pp$": Fully coupled-channel complex scaling method which is our method to treat the ``$K^-pp$", and a chiral SU(3)-based $\bar{K}N$(-$\pi Y$) potential that we have employed. We give an additional explanation on the treatment of this $\bar{K}N$ potential, since it is an energy-dependent potential. Here, we remark that the ``$K^-pp$" is considered also in the present study. In other words, as the $K^-pp$ system, we consider a $\bar{K}NN$-$\pi\Sigma N$-$\pi\Lambda N$ coupled-channel system with $J^\pi=0^-$ and $T=1/2$, similarly to the past studies as mentioned in the previous section.  

\subsection{Fully coupled-channel complex scaling method}

First, we consider the two-body case, which corresponds to the $\Lambda(1405)$. Since the Dalitz's  speculation that the excited hyperon $\Lambda(1405)$ would be a $\bar{K}N$ molecule state \cite{Dalitz:KbarN, Dalitz:L1405}, there have been many studies based on this picture. Most of these past studies suggest the importance of the coupling of $\bar{K}N$ and $\pi\Sigma$ in the $\Lambda(1405)$ \cite{ChU:Review}. Such a channel coupling is expected to be important also in the three-body case of the $K^-pp$ system, if it forms a (quasi-)bound state. In the $K^-pp$ case, the coupling of $\bar{K}N$ to $\pi\Lambda$ is involved as well as that to $\pi\Sigma$ in terms of isospin. Therefore, in our study of the $K^-pp$ system, we should consider a coupled-channel state involving $\bar{K}NN$, $\pi\Sigma N$ and $\pi\Lambda N$ components. Second, according to most of theoretical studies in past, the binding energy of the ``$K^-pp$" should be less than 100 MeV, as mentioned in the introduction. This means that the ``$K^-pp$" is not completely bound but exists as a resonant state between the $\bar{K}NN$ and $\pi\Sigma N$ thresholds. (It is noted that the $\pi\Sigma N$ threshold is energetically 103 MeV below the $\bar{K}NN$ threshold.) 

From these facts, the following two ingredients are considered to be key factors for theoretical investigation of the $K^-pp$ system: {\bf 1. coupled channel} and {\bf 2. resonance}. We have developed ``fully coupled-channel complex scaling method (Full ccCSM)" which can deal with both ingredients adequately \cite{Full-ccCSM_AY:Dote}. 

As for the first ingredient, {\it coupled-channel aspect}, all channels which can couple to the ``$K^-pp$" state, $\bar{K}NN$, $\pi\Sigma N$, and $\pi\Lambda N$, are explicitly treated in our calculation. In the actual calculation, the ``$K^-pp$" wave function is expressed as 
\begin{eqnarray}
|\Phi_{``K^-pp"}\rangle \quad = \quad  \sum_{ch=1}^8 \sum_{n=1}^N \; C^{(ch)}_n \, F^{(ch)}_n (\bm{x}_1, \bm{x}_2) \; 
|S_{B_1 B_2 (ch)}=0\rangle  \, \times \,|(M B_1 B_2)_{(ch)}; T=1/2, T_z=1/2 \rangle. \label{Kpp-wfn}
\end{eqnarray}
Here, the last term of the above equation is the isospin-flavor wave function, in which the label $(M B_1 B_2)_{(ch)}$ indicates the channels of $\bar{K}NN$, $\pi\Sigma N$, and $\pi\Lambda N$. It is noted that there are eight channels ($ch=1, \ldots, 8$), as a result of the anti-symmetrization for baryons $B_1$ and $B_2$. (See Table I of Ref. \cite{Full-ccCSM_AY:Dote}.) The spatial wave function $F^{(ch)}_n (\bm{x}_1, \bm{x}_2)$ is expanded with the correlated Gaussian basis functions \cite{CG:Suzuki} in which three types of Jacobi-coordinate sets $\{\bm{x}_1, \bm{x}_2\}$ are taken into account. Similarly to all earlier studies \cite{StrangenessSummary_2016}, spin of baryons in each channel ($S_{B_1 B_2 (ch)}$) is assumed to be zero. The complex parameters $\{ C^{(ch)}_n \}$ are determined by diagonalizing the complex-scaled Hamiltonian matrix as explained latter.

For the second ingredient, {\it resonance treatment}, we have relied on the complex scaling method (CSM), which has succeeded in many studies of resonant states of stable/unstable nuclei \cite{CSM:Myo2}. Here, we give a brief explanation of the CSM. In the CSM, Hamiltonian and wave functions are transformed with complex rotation (or complex scaling): 
\begin{equation}
\bm{r}_j \rightarrow \bm{r}_j \exp(i\theta) \quad {\rm and} \quad \bm{p}_j \rightarrow \bm{p}_j \exp(-i\theta),
\end{equation} 
where $\bm{r}_j$ and $\bm{p}_j$ are spatial coordinates of the system and their conjugate momenta, respectively. The real variable $\theta$ is called a scaling angle. By such a complex rotation, resonant wave functions, which diverge in asymptotic region, are transformed to damping functions with oscillation. Therefore, resonant wave functions can be represented with $L^2$-integrable functions such as Gaussian basis functions, as well as bound-state wave functions. Furthermore, there is an important theorem in the CSM, which is called {\it ABC theorem}. In case of non-relativistic kinematics, the energy of continuum states ($E_c$) varies as $E_c^\theta=E_c e^{-2i\theta}$ with scaling angle $\theta$, as is easily noticed by considering the fact that the kinetic energy is given as a form of $\bm{p}^2_j / 2\mu_j$. ($\mu_j$ is a reduced mass with respect to the $j$-th coordinate.) In other words, the energy of continuum states depends on the $\theta$ value, and it is found that their energies appear along  a line satisfying $\tan^{-1}({\rm Im} \, E_c^\theta \, / \, {\rm Re} \, E_c^\theta) = -2\theta$ on a complex energy plane. This line is called {\it $2\theta$ line}. On the other hand, the energy of resonant states as well as bound states are independent of the $\theta$ variation, according to the ABC theorem. With the help of this nature, we can separate resonant states from continuum states. In our actual calculations, we carry out the diagonalization of the Hamiltonian with finite numbers of Gaussian basis functions to obtain eigenvalues of resonant states. Although many eigenvalues are obtained, those indicating resonant states should appear off the $2\theta$ line on the complex energy plane, and they do not move even if the scaling angle $\theta$ is varied. Therefore, we can identify resonant states. At the end of this paragraph, we summarize advantageous points of the CSM. In the CSM,  resonant states are correctly treated; Resonant states are treated as Gamow states with complex energy, and they are obtained under the correct boundary condition (outgoing wave condition). In a technical viewpoint, the CSM is a convenient method because it can treat resonant states almost in the same way as bound states. As mentioned above, the CSM matches well with the diagonalization method with Gaussian basis function, which is often used to study bound states. In addition, the CSM provides us with wave functions of resonant states. By using them, we can investigate properties of these resonant states. It should be noted  that this method can be applied to many-body systems beyond two-body systems straightforwardly.

\subsection{Chiral SU(3)-based $\bar{K}N$(-$\pi Y$) potential}

In the present fully coupled-channel calculation, pion as well as antikaon are explicitly involved in the ``$K^-pp$". These mesons are so-called Nambu-Goldstone bosons, which are associated with the spontaneous breaking of chiral symmetry which the quantum chromodynamics (QCD) possesses at vanishing quark masses. Nature of these bosons should be governed by chiral dynamics. In our previous study \cite{Full-ccCSM_AY:Dote}, we employed a phenomenological $\bar{K}N$ potential \cite{AY_2002}. On the other hand, since the chiral theory is an effective theory of QCD which is the fundamental theory for hadron physics, we also employ a chiral SU(3)-based $\bar{K}N$ potential in order to investigate the $K^-pp$ system as theoretically as possible. In our early study \cite{ccCSM-KN_NPA}, we proposed such a $\bar{K}N$(-$\pi Y$) potential, which was derived from the chiral effective Lagrangian, similarly to the coupled-channel chiral dynamics \cite{ChU:KSW} or the chiral unitary model \cite{ChU:OR}. It is noted that we take into account only Weinberg-Tomozawa term of the Lagrangian and Gaussian form factors in our potential. In the present study, we use the simplest version of our non-relativistic potential (referred to as NRv2c potential in the original paper \cite{ccCSM-KN_NPA}). The potential between channels $\alpha$ and $\beta$ with isospin $I$ is given as  
\begin{equation}
\hat{V}^{MB,I}_{\alpha,\beta} \; = \; -\frac{C^I_{\alpha\beta}}{8f^2_\pi} \, \left( \omega_\alpha+\omega_\beta \right) \, \sqrt{\frac{1}{m_\alpha m_\beta}} \; g^I_{\alpha\beta} (r) \, | \alpha \rangle \langle \beta | \quad {\rm with} \quad g^I_{\alpha\beta}(r) \; =\;  {\cal N}^I_{\alpha\beta} \exp\left\{-\left(\frac{r}{d^I_{\alpha\beta}}\right)^2\right\}, \label{chiral_pot}
\end{equation}
where the channels $\alpha$ and $\beta$ are $\bar{K}N$, $\pi\Sigma$ or $\pi\Lambda$, and the isospin $I=0$ or $1$. In the above equation, $f_\pi$ is the pion decay constant, $C^I_{\alpha\beta}$ is Clebsh-Gordan coefficient of SU(3) algebra, and $m_\alpha$ is a meson mass in a channel $\alpha$. The function $g_{\alpha\beta} (r)$ is a normalized Gaussian function with the normalization factor ${\cal N}^I_{\alpha\beta}$, where $r$ is the inter-particle distance. It should be remarked that $\omega_{\alpha(\beta)}$ is a meson energy in the channel $\alpha$ ($\beta$), which is attributed to the chiral dynamics. More details of our potential are given in the original paper \cite{ccCSM-KN_NPA}. 

In our $\bar{K}N$(-$\pi Y$) potential, the pion decay constant $f_\pi$ is treated as a parameter, and it is varied from 90 MeV to 120 MeV so as to cover the experimental values of pion and kaon decay constants. For each $f_\pi$ value, the range parameters of the normalized Gaussian function, $d^I_{\alpha\beta}$, are adjusted to reproduce the experimental value of $\bar{K}N$ scattering length. In the original study, we referred to the old value of $\bar{K}N$ scattering length, which was obtained by Martin's analysis \cite{Exp:ADMartin}. By the way, the $1s$ level energy shift of kaonic hydrogen atom was precisely measured by SIDDHARTA collaboration \cite{Exp:SIDDHARTA}, and the $\bar{K}N$ scattering length is deduced from this data by the analysis using a coupled-channel chiral dynamics \cite{ChU:IHW}. In the present study, we have utilized this latest value as well, to construct the $\bar{K}N$(-$\pi Y$) potential. Hereafter, the original version of our potential, which is constrained with the old $\bar{K}N$ scattering length is denoted as {\it Chiral-old potential}, whereas its updated version, which is constrained with the latest $\bar{K}N$ scattering length, is denoted as {\it Chiral-latest potential}.

The non-relativistic Hamiltonian, which includes the above-mentioned $\bar{K}N$(-$\pi Y$) potential, is as follows: $\hat{H}_{``K^-pp"} = \hat{M} + \hat{T} + \hat{V}_{NN} + \hat{V}_{MB}$, which is composed of the mass term $\hat{M}$, the kinetic term $\hat{T}$, a nucleon-nucleon potential $\hat{V}_{NN}$ and a meson-baryon potential $\hat{V}_{MB}$. Since all channels are explicitly treated in the present calculation, we need the mass term, which gives the threshold mass for $\bar{K}NN$, $\pi\Sigma N$ and $\pi\Lambda N$ channels; $\hat{M} = \sum_\alpha (m_\alpha + M_{1, \alpha}+M_{2, \alpha}) |\alpha\rangle \langle \alpha |$, where $m_\alpha$ and $M_{1(2), \alpha}$ are the mass of a meson and a baryon in a channel $\alpha$, respectively. The term $\hat{T}$ is a non-relativistic kinetic-energy operator with respect to Jacobi coordinates $\{ \bm{x}_1, \bm{x}_2 \}$. In the current study, we employ the Argonne v18 realistic $NN$ potential \cite{Av18} as a $NN$ potential $\hat{V}_{NN}$, and our $\bar{K}N$(-$\pi Y$) potential as a $MB$ potential $\hat{V}_{MB}$. It is noted that $YN$ and $\pi N$ potentials are not considered, since their contributions might be minor compared to $NN$ and $\bar{K}N$ potentials. 

\subsection{Treatment of energy-dependent potential}

As shown in Eq. (\ref{chiral_pot}), the $\bar{K}N$(-$\pi Y$) potential that is used in the current study has an energy dependence, since it includes explicitly the meson energy $\omega_{\alpha \, (\beta)}$. It is not so simple to handle such an energy-dependent potential: The strength of the potential is determined by the meson energy (or meson-baryon energy, which means the energy of a meson-baryon pair). However, to know the meson energy, we need the wave function of the system which is calculated using the potential. Therefore, we have conducted self-consistent calculations on the meson-baryon energy. In the current study, such energy-dependent potentials are dealt with, based on a prescription proposed in an earlier study with a variational method in a single $\bar{K}NN$ channel \cite{Kpp:DHW}.     

Here, we explain how to treat the energy-dependent potential briefly. Essentially, we have extended the previous prescription for a single channel calculation \cite{Kpp:DHW} to a coupled-channel calculation, by averaging the meson-baryon energy and the mass threshold with the amplitude of the wave function of each channel \cite{Full-ccCSM_Chiral:Dote}. First, we consider the meson binding energy which is defined as the difference between the energy of the total ``$K^-pp$" system and that of its baryon-part system, corresponding to the antikaon binding energy in the earlier single-channel study. The operator of the meson binding energy $\hat{B}_M$ is defined as 
\begin{equation}
\hat{B}_M = -(\hat{H}_{``K^-pp"} - \hat{H}_{BB} - \hat{m}_{M}), 
\end{equation}
where the terms $\hat{H}_{``K^-pp"}$, $\hat{H}_{BB}$ and $\hat{m}_{M}$, are operators of the total Hamiltonian, the baryon-part Hamiltonian and the meson mass, respectively. It is noted that the last term $\hat{m}_{M}$ is needed, because it is included in the total Hamiltonian but not so in the baryon-part Hamiltonian; $\hat{H}_{BB} = \hat{M}_{BB} + \hat{T}_{BB} + \hat{V}_{NN}$, where $\hat{M}_{BB}$ and $\hat{T}_{BB}$ indicate the mass and kinetic energy operators of the baryon-baryon system, respectively. Next, using this $\hat{B}_M$ and the ``$K^-pp$" wave function $\Phi_{``K^-pp"}$, the energy of a meson-baryon pair (meson-baryon energy, $MB$ energy), $\sqrt{s}_{MB}$, is estimated in two ways based on extreme two pictures, {\it field picture} and {\it particle picture}, in the same manner as the earlier study \cite{Kpp:DHW}:
\begin{equation}
\sqrt{s}_{MB} = \left\{ 
\begin{array}{lclcl}
\langle \Phi_{``K^-pp"}| & \hat{M}_B+\hat{m}_M - \hat{B}_M  & |\Phi_{``K^-pp"} \rangle & \ldots & {\rm Field \; picture},\\
\langle \Phi_{``K^-pp"}| & \hat{M}_B+\hat{m}_M - \hat{B}_M/2 & |\Phi_{``K^-pp"}\rangle & \ldots & {\rm Particle \; picture}.
\end{array}
\right. 
\end{equation}
We make a comment on this prescription for the estimation of the $MB$ energy. The $MB$ energy is the energy of a two-body system in the total three-body system. Since such a subsystem energy is not an eigenvalue of the total Hamiltonian, it can not be determined uniquely in principle. Therefore, we examine the two extreme pictures to estimate the $MB$ energy as explained above. In the field picture, the meson binding energy is fully carried by a single meson-baryon bond, while only its half is carried in the particle picture. 

For the $MB$ energy, $\sqrt{s}_{MB}$, we perform self-consistent calculations. First, we assume a value $\sqrt{s}_{MB, In}$ to determine the strength of the meson-baryon potential. With the Hamiltonian including that potential, we search for a resonance pole of ``$K^-pp$" by full ccCSM calculation. Then, we evaluate the $MB$ energy (denoted as $\sqrt{s}_{MB, Out}$) with the wave function of the found resonant state, with the above-mentioned prescription. We repeat such calculations with various values of the assumed $MB$ energy $\sqrt{s}_{MB, In}$, comparing it to the estimated $MB$ energy $\sqrt{s}_{MB, Out}$. When we find a solution satisfying a condition $\sqrt{s}_{MB, In}=\sqrt{s}_{MB, Out}$, it is a self-consistent solution which should be obtained. It is noted that the $MB$ energy is a complex value as well as the energy of resonant states which is treated as a complex energy of a Gamow state in the current study. The self-consistent calculation is carried out with a complex-valued $MB$ energy. 

\begin{figure}[t]
  \centerline{
\includegraphics[width=200pt]{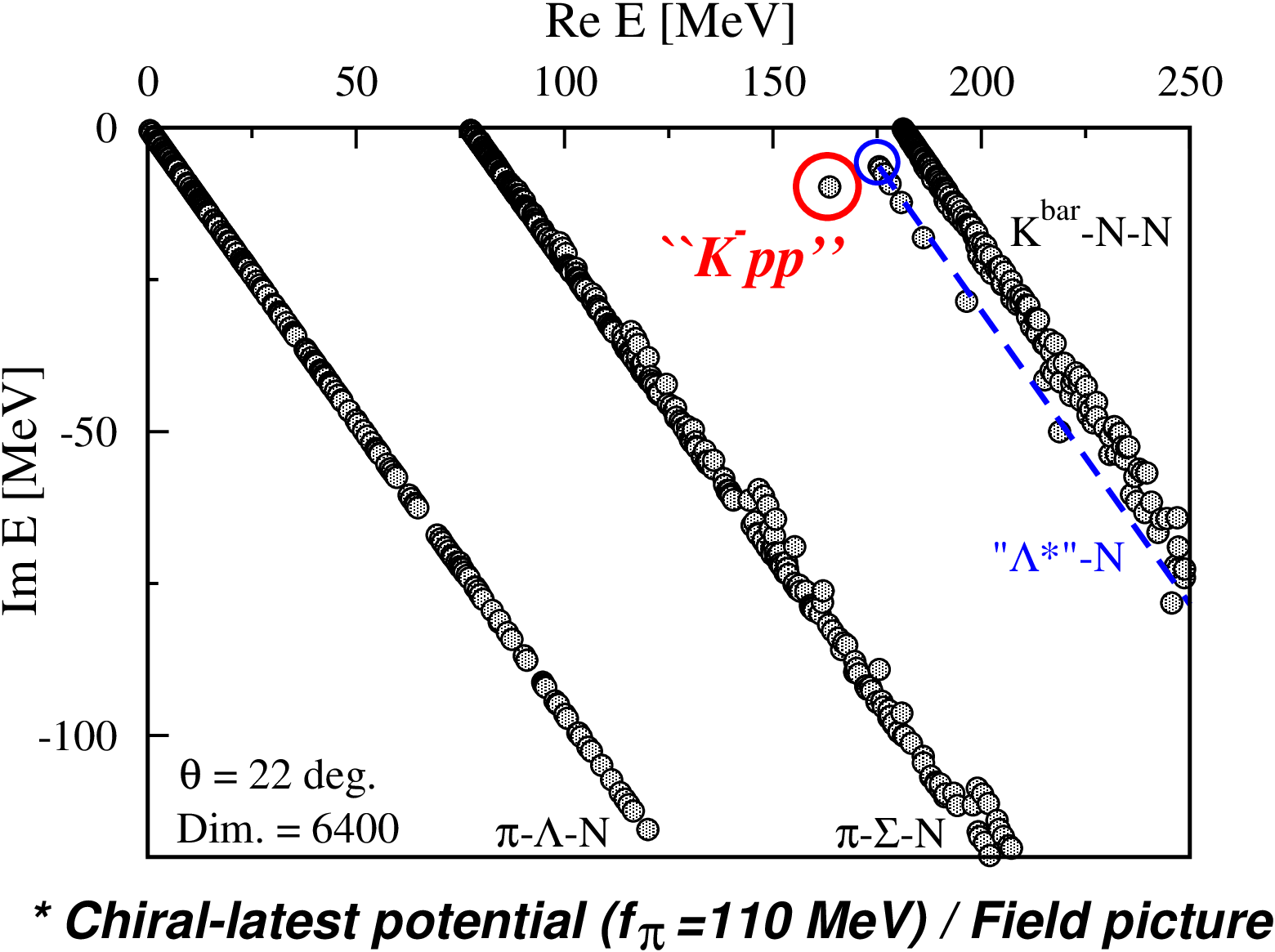}
}
  \caption{(Color online)  Eigenvalue distribution on the complex energy plane, where Chiral-latest potential with $f_\pi=110$ MeV is used and the field picture is employed. The eigenvalue marked with a red circle corresponds to the ``$K^-pp$" resonance. The eigenvalue marked with a blue circle corresponds to the $\Lambda^*$ resonance. (See the text.) The thresholds of $\pi\Lambda N$, $\pi\Sigma N$ and $\bar{K}NN$ are located at 0 MeV, 77 MeV, and 181 MeV on the real axis, respectively. \label{Eigenvalues_Kpp-Latest-fp110-F}}
\end{figure}

\section{RESULTS AND DISCUSSION}

\subsection{Results}

Here, we show the results of ``$K^-pp$" calculated with Full ccCSM using our chiral SU(3)-based potential. The setup of the calculation is as follows: For each Jacobi coordinate in each channel, 20 Gaussian basis functions are used to expand the spacial part of the ``$K^-pp$" wave function, whose width parameters range from 0.1 fm to 20 fm in geometrical progression. Taking into account the number of sets of the Jacobi-coordinates and that of channels, the size of the complex-scaled Hamiltonian matrix amounts to 6400 dimension. Such a large-size matrix is diagonalized directly without any approximation. The scaling angle $\theta$ is chosen to be 22 degree in all calculations, because we have confirmed that physical quantities such as norm of each channel and inter-particle distances, in addition to the resonance energy, are the most stable at this angle. Both of our $\bar{K}N$(-$\pi Y$) potentials, Chiral-old and Chiral-latest potentials, are utilized. As for the self-consistency for the meson-baryon energy, both ansatzes of field picture and particle picture are examined.  

Figure \ref{Eigenvalues_Kpp-Latest-fp110-F} represents the distribution of the complex eigenvalues when the self-consistent solution of the ``$K^-pp$'' resonance is obtained. Here, as a typical case, Chiral-latest potential with $f_\pi=110$ MeV is used and the field pictures employed. There are many eigenvalues plotted on the figure. We can interpret them as follows: First, it is found that there are three groups of eigenvalues appearing along lines which start from the $\pi\Lambda N$, $\pi\Sigma N$, and $\bar{K}NN$ thresholds. These eigenvalues correspond to the three-body continuum states of $\pi\Lambda N$, $\pi\Sigma N$, and $\bar{K}NN$, respectively, since they are along the $2\theta$ lines starting from the corresponding thresholds. Second, the eigenvalue marked with a blue circle in the figure corresponds to the $\Lambda^*$ resonance, which is a resonant state of the two-body system of the $\bar{K}N$-$\pi\Sigma$ with isospin zero. Therefore, a group of eigenvalues appearing along the $2\theta$ line which starts from the $\Lambda^*$ indicates the two-body continuum states of this $\Lambda^*$ and a nucleon. At last, there is the eigenvalue isolated from these four lines. It is the eigenvalue of the ``$K^-pp$'' three-body resonance. Thus, we can confirm that Full ccCSM works correctly in the $K^-pp$ study.  

\begin{figure}[t]
  \centerline{
\includegraphics[width=200pt]{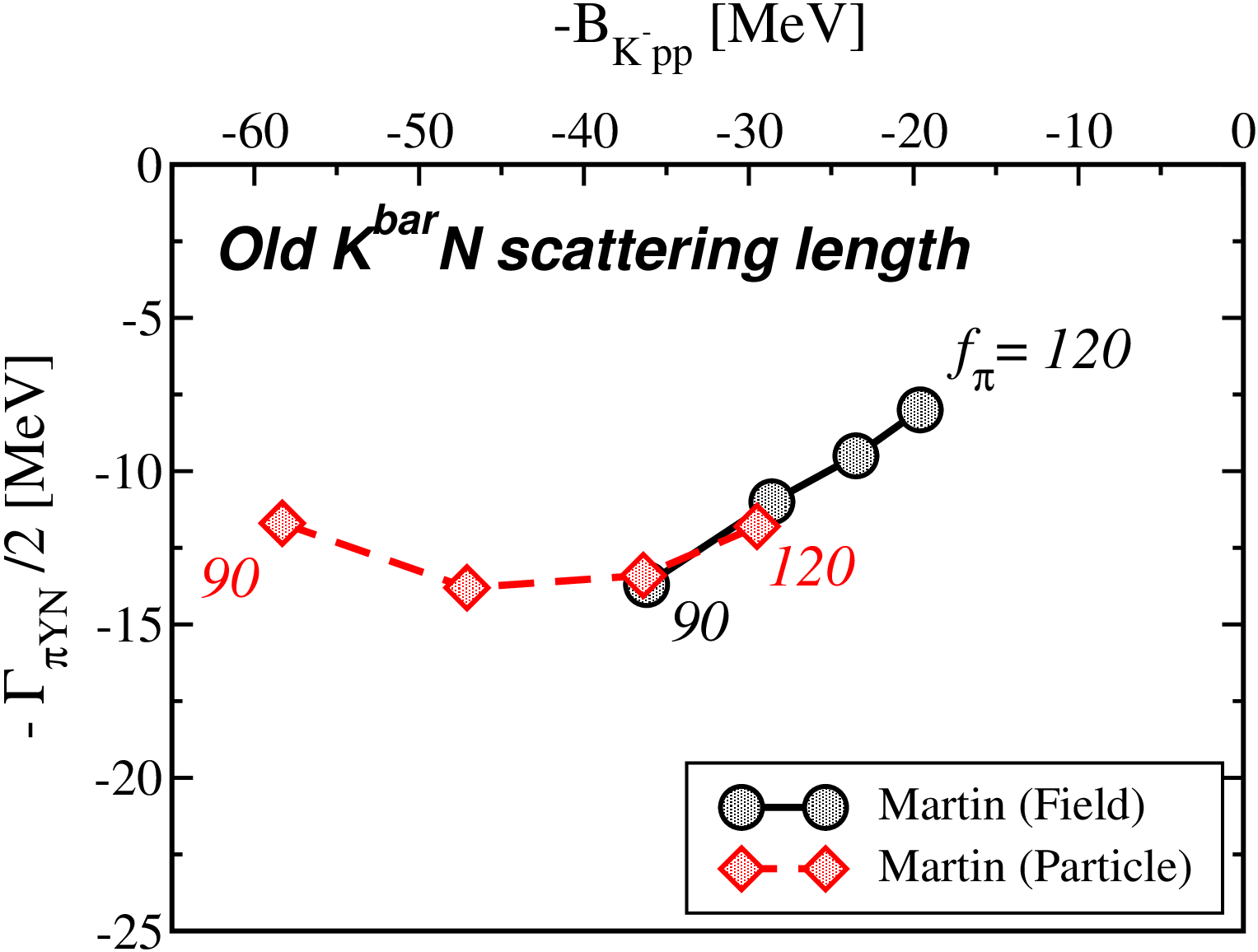}
\hspace{1cm}
\includegraphics[width=200pt]{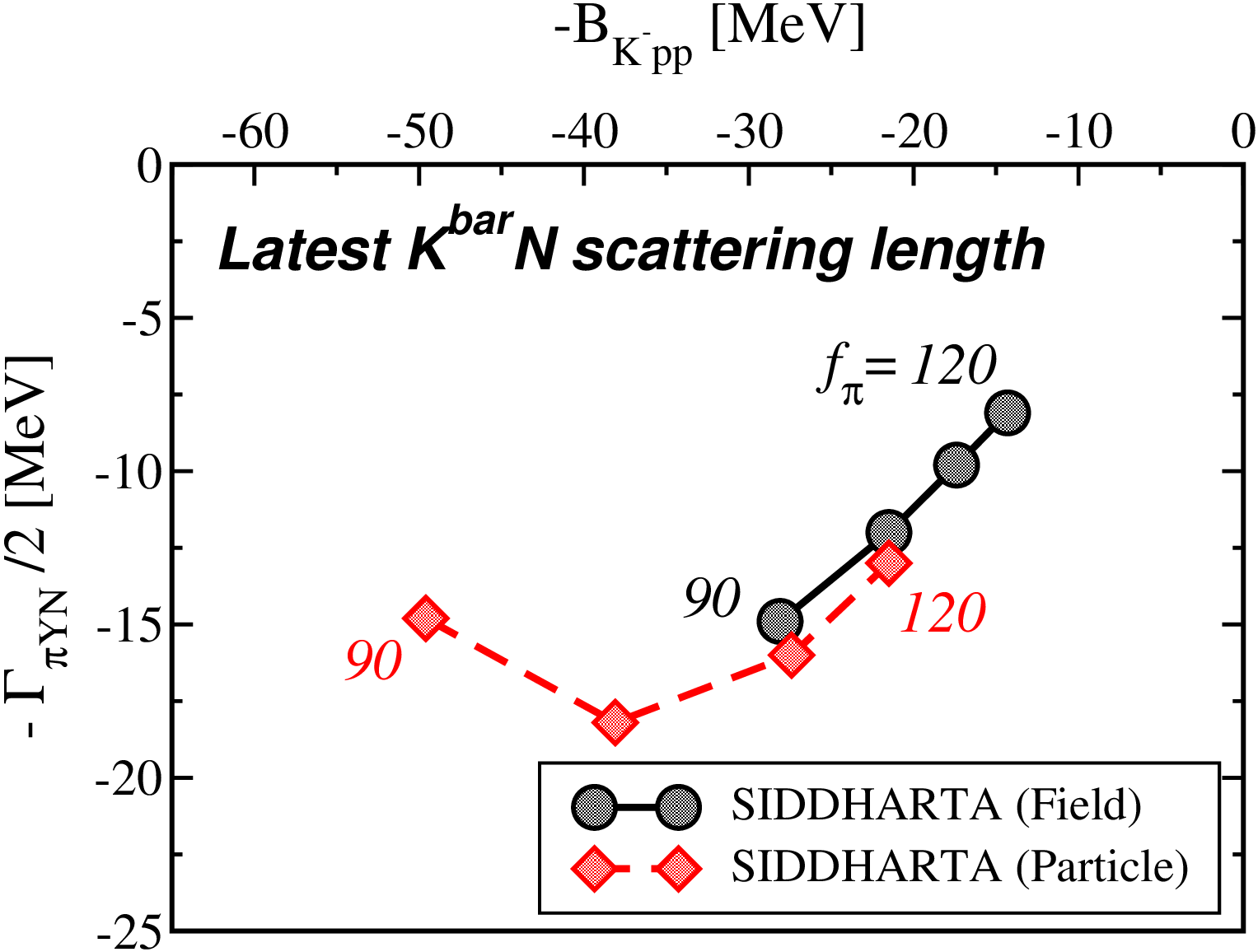}
}
  \caption{(Color online) Binding energy ($B_{K^-pp}$) and mesonic decay width ($\Gamma_{\pi YN}$)  of the ``$K^-pp$" system. The left (right) panel shows the result calculated with Chiral-latest (Chiral-old) potential, which is constraint with the latest (old) $\bar{K}N$ scattering length. In each panel, symbols of black circles (red diamonds) indicate that the field picture (particle picture) is employed as the ansatz for the self-consistency. The value of $f_\pi$, which is a parameter of our potentials (Eq. (\ref{chiral_pot})), is varied with 10-MeV interval. It should be remarked that the potentials are tuned to reproduce the $\bar{K}N$ scattering length with each $f_\pi$ value. All energies are given in unit of MeV. \label{Result_Kpp-pole}}
\end{figure}

Figure \ref{Result_Kpp-pole} shows all results of Full ccCSM calculation of the ``$K^-pp$" system with our chiral SU(3)-based potentials. In all examined cases, two types of $\bar{K}N$(-$\pi Y$) potentials with various $f_\pi$ values (90 to 120 MeV) and two ansatzes for the self-consistent calculation, we have successfully found self-consistent solutions of the ``$K^-pp$" resonance, as their binding energy and decay with are plotted in the figure. As shown in Fig. \ref{Result_Kpp-pole}, the results obtained with Chiral-old and Chiral-latest potentials are found to have similar dependence on the parameter $f_\pi$ and the ansatz for the self-consistency. In both potentials, the field picture gives small binding energy and small decay width, compared with the particle picture. In particular, in case of the field picture a linear correlation is found between the binding energy and the decay width. When the results with two potentials are compared, the ``$K^-pp$" calculated with Chiral-latest potential is more shallowly bound than with Chiral-old potential. This is attributed to the fact that the latest potential is less attractive than the old potential. Indeed, when the two-body system of the $\bar{K}N$-$\pi\Sigma$ system with isospin zero is investigated, its resonant state, which corresponds to the $\Lambda(1405)$, is obtained more shallowly bound with Chiral-latest potential than with Chiral-old potential: $(B_{\bar{K}N}, \Gamma/2)=(8.0, 13.4)$ MeV in case of Chiral-latest potential, whereas it is found at $(B_{\bar{K}N}, \Gamma/2)=(17.2, 16.6)$ MeV in case of Chiral-old potential. (Here, $B_{\bar{K}N}$ and $\Gamma$ represent the binding energy measured from $\bar{K}N$ threshold and decay width of the resonant state, respectively. The potential parameter $f_\pi$ is set to be 110 MeV.)  As mentioned in the introduction, the $I=0$ $\bar{K}N$ potential is strongly attractive and it is expected to give major contribution to the binding of the ``$K^-pp$". The difference of the binding energy is about 10 MeV between both potential cases. In case of Chiral-latest potential, which is constrained with SIDDHARTA data, the binding energy and the mesonic decay width are obtained to be 
\begin{equation}
(B_{K^-pp}, \; \Gamma_{\pi YN}/2) \quad = \quad \left\{
\begin{array}{llcl}
(14-28, \; 8-15) & {\rm MeV} & \ldots & {\rm Field \; picture},\\
(21-50,\; 13-19) & {\rm MeV} & \ldots & {\rm Particle \; picture}.
\end{array}
\right .
\end{equation}

\begin{table}[b]
\caption{Details of the ``$K^-pp$" resonance. The latest version of our chiral SU(3)-based potential with $f_\pi=110$ MeV is used. $B_{K^-pp}$ and $\Gamma_{\pi YN}$ are the binding energy and mesonic decay width of the ``$K^-pp$" system, respectively. $R_{NN}$ is a mean distance between two nucleons in the ``$K^-pp$" resonance. ${\cal N}(\bar{K}NN)$ is a norm of the $\bar{K}NN$ component in the ``$K^-pp$" resonance, and so on. The top and middle lines (``Chiral-latest / Field" and ``Chiral-latest / Particle") show the results with field picture and particle picture, respectively. On the bottom line (``Phenomenological"), the result obtained with Full ccCSM employing a phenomenological potential \cite{AY_2002} is shown as a reference. The units of energy and length are MeV and fm, respectively.}
\label{Kpp-detail}
\tabcolsep6pt\begin{tabular}{lcccccc}
\hline
  & $B_{K^-pp}$  & $\Gamma_{\pi YN}/2$  & $R_{NN}$  & ${\cal N}(\bar{K}NN)$ & ${\cal N}(\pi\Sigma N)$ &  ${\cal N}(\pi\Lambda N)$ \\
\hline
Chiral-latest / Field    & 17.4 & 9.8  & $2.14-i0.16$  & $1.131-i0.045$ & $-0.128+i0.041$ & $-0.002+i0.004$
\\
Chiral-latest / Particle & 27.3 & 15.9 & $1.80-i0.06$  & $1.286-i0.110$ & $-0.283+i0.103$ & $-0.002+i0.008$ \\
Phenomenological\tabnoteref{AY-case}                            & 51   & 16   & $1.86+i0.14$  & $1.004-i0.286$ & $-0.002+i0.276$ & $-0.002+i0.010$ \\
\hline
\end{tabular}
\tablenote[AY-case]{This is quoted from Ref. \cite{Full-ccCSM_AY:Dote}}
\end{table}

Details of the obtained ``$K^-pp$" resonance are shown in TABLE \ref{Kpp-detail}, in which a parameter in our chiral potential, $f_\pi$, is set to be 110 MeV as a typical case. As mentioned above, the ``$K^-pp$" calculated with the field picture (top line) is shallowly bound, compared with the particle picture (bottom line). Accordingly, the size of the system is larger in the field picture than in the particle picture, as indicated by the mean distance between two nucleons, $R_{NN}$. Here, we note on physical quantities of resonant states. They are obtained as complex values in our treatment, because resonant states are treated as Gamow states. However, when their imaginary part is small, it is expected that these complex-valued quantities can be interpreted as usual, similarly to the case that they are real values \cite{CSM-meaning:Berggren, ChU:EMsize}. The norm of each component of the ``$K^-pp$" resonance is also obtained as a complex value, as given in columns of ${\cal N}(\bar{K}NN)$, ${\cal N}(\pi\Sigma N)$ and  ${\cal N}(\pi\Lambda N)$ in the table. Although these values are complex, we can know the composition of the ``$K^-pp$" resonance clearly, considering magnitudes of the norms: The $\bar{K}NN$ component is dominated there, the $\pi\Sigma N$ is slightly involved,  and the $\pi\Lambda N$ is very marginal. Such a composition is similar to the case investigated with a phenomenological potential in our previous study \cite{Full-ccCSM_AY:Dote}.

\subsection{Discussion}

In this section, we make some discussions, based on the results of our present calculations with Full ccCSM using our chiral SU(3)-based potentials. 

As mentioned in the introduction, one of important issues in kaonic nuclei is the question whether kaonic nuclei can form a dense state or not. The particle number of the ``$K^-pp$" system that we have investigated is too few to consider the ``density". However, the mean distance between two nucleons are expected to give some hint to consider this question, because we can estimate it in case of the nuclear matter. Indeed, the mean $NN$ distance is known to be 2.2 fm in case of the nuclear matter with the normal nuclear density ($\rho_0=0.16$ fm$^{-3}$). Compared with this value, the ``$K^-pp$" obtained with the field picture is found to correspond to a matter with the normal density, since its $NN$ distance is close to that of the normal nuclear matter: $R_{NN}=2.1$ fm as shown in TABLE \ref{Kpp-detail}. With the particle picture, the $NN$ distance indicates a more dense state. We comment on the case that a phenomenological potential \cite{AY_2002} is employed. In this case, the ``$K^-pp$" resonance is obtained as a rather deeply bound state and a compact system with $R_{NN}=1.9$ fm, as shown on the bottom line of the table. This mean $NN$ distance corresponds to $\sim 1.6 \rho_0$ which indicates somewhat dense matter. Therefore, the answer to the above-mentioned question is dependent on a type of $\bar{K}N$(-$\pi Y$) potential: If the potential in the real world is weakly attractive like chiral SU(3)-based potential (and the field picture is rather favored), we should stay at the normal matter world, even if an antikaon gives an attraction to the nuclear matter. On the other hand, if the potential is so strongly attractive as suggested by a phenomenological potential, we can go to a dense matter world with antikaons. In other words, if this is the case, kaonic nuclei could be a doorway to the dense matter. 

\begin{figure}[t]
  \centerline{
\includegraphics[width=300pt]{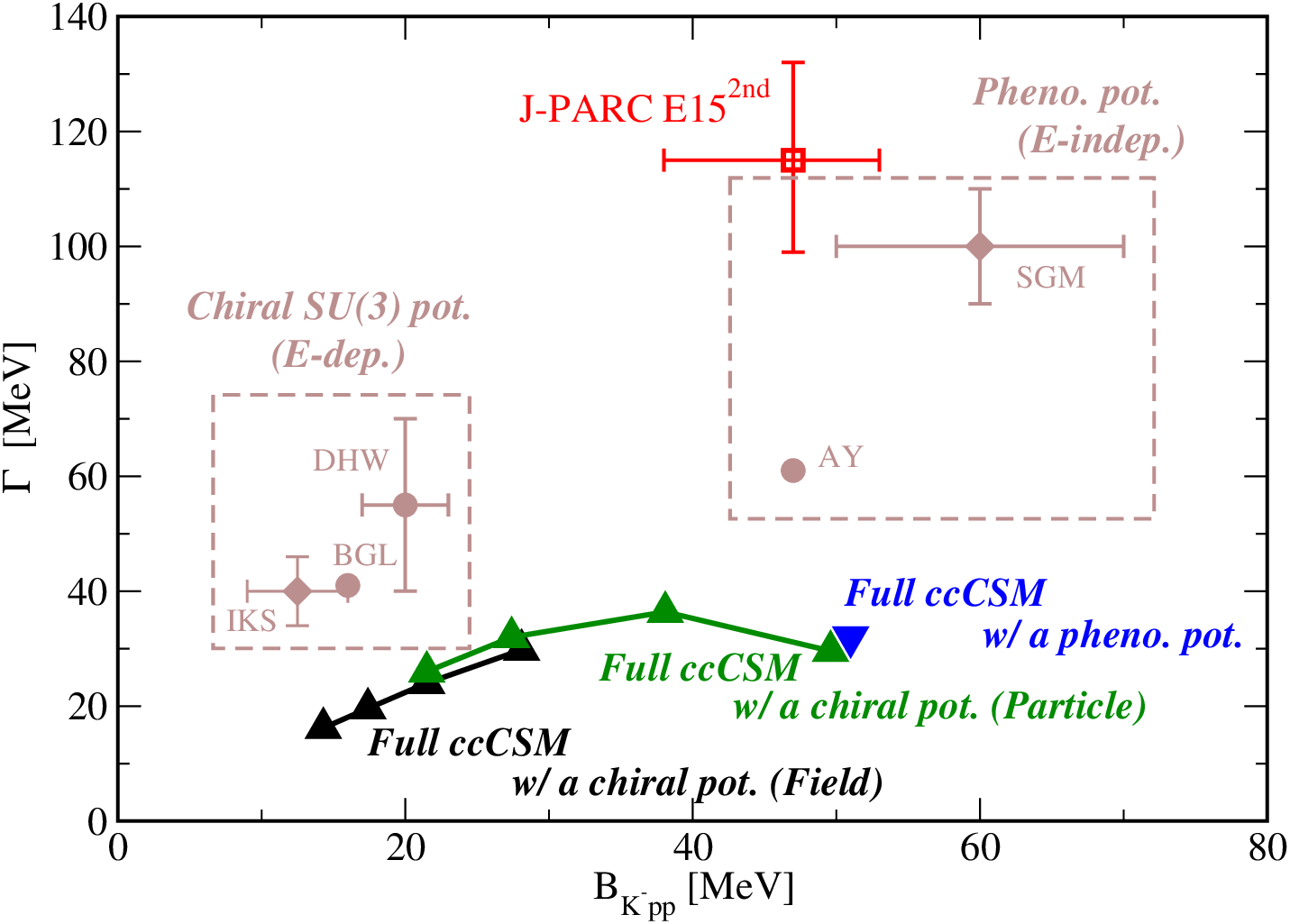}
}
  \caption{(Color online) Summary of the latest $K^-pp$ studies. The results of our current studies and J-PARC E15$^{\rm 2nd}$ are shown together with those of earlier theoretical studies ({\it AY} \cite{Kpp:AY}, {\it SGM} \cite{Faddeev:Shevchenko}, {\it DHW} \cite{Kpp:DHW}, {\it BGL} \cite{Kpp:BGL} and {\it IKS} \cite{Kpp:IKS}, which are plotted with brown symbols). As for our study with {\it Full ccCSM} using a chiral SU(3)-based potential, which is constrained by the SIDDHARTA data, results obtained with both ansatzes for the self consistency are shown: Field picture (Particle picture) is plotted with symbols of black (green) up-pointing triangles, which is denoted as {\it w/ a chiral pot. (Field (Particle))}. In addition, the  result of Full ccCSM employing a phenomenological potential is plotted with a symbol of a blue down-pointing triangle, which is denoted as {\it w/ a pheno. pot.}. The symbol of a red square represents the result of J-PARC E15$^{\rm 2nd}$ experiment, which is plotted with a interpretation that the observed signal is a $K^-pp$ quasi-bound state. It is noted that results of other earlier experiments (FINUDA \cite{Kpp:exp_FINUDA}, DISTO \cite{Kpp:exp_DISTO} and J-PARC E27 \cite{Kpp-ex:JPARC-E27}) can not be plotted on this figure because the reported binding energy is more than 90 MeV, if they are interpreted as a $K^-pp$ quasi-bound state. \label{Summary_Latest-Kpp}}
\end{figure}

Figure \ref{Summary_Latest-Kpp} is the updated summary of various theoretical and experimental studies of the $K^-pp$ system, which includes our results as reported above and the latest result of J-PARC E15 experiment (the second run of their experiment, J-PARC E15$^{\rm 2nd}$). Calculated and reported binding energy and decay width of the $K^-pp$ system are plotted in the figure. We remark on the comparison of theoretical results and experimental results. The comparison with respect to pole positions of resonant states is not so simple, in particular, when the resonant state involves a large decay width. In such a case, it is well known that the peak position of the experimentally observed spectrum (cross section) does not directly correspond to the pole position of the resonance, as pointed out in early studies of $\Sigma$ hypernuclei \cite{Morimatsu-Yazaki}. Here, we make such a simple comparison just as a trial, though we know the above-mentioned property of the peak position. In addition, we make a naive assumption that all experimental signals are due to the $K^-pp$ quasi-bound state. Based on this assumption, it is obvious that our results as well as other theoretical results can not explain the earlier experimental results, FINUDA, DISTO and J-PARC E27, which indicate a deeply-bound state with $\sim$100 MeV binding energy. On the other hand, the J-PARC E15$^{\rm 2nd}$ result, especially its binding energy, is among the theoretical predictions. As mentioned in the introduction, according to earlier theoretical studies, chiral-type energy dependent potentials make the ``$K^-pp$" shallowly bound ($B_{K^-pp} \sim 20$ MeV), whereas phenomenological energy-independent potentials make it rather deeply bound ($B_{K^-pp} > 50$ MeV). Compared with these theoretical results, the binding energy of the J-PARC E15$^{\rm 2nd}$ result is found to be close to predictions with phenomenological potentials rather than those with chiral-type potentials. As for the current result of Full ccCSM calculation with the latest version of our potential, when the field picture is employed for the self-consistency of $MB$ energy, the result is similar to those of previous calculations with chiral-type potentials, and the calculated binding energy does not reach to that of the J-PARC E15$^{\rm 2nd}$ result, which is 30 MeV at most. However, when the particle picture is employed, the calculated binding energy increases up to 50 MeV. This result indicates that there is a room to explain the latest J-PARC E15 result also with chiral-type potentials. By the way, there is a large discrepancy in the decay width between our result and the J-PARC E15$^{\rm 2nd}$ result. The reason of this discrepancy has been considered to be attributed to the non-mesonic decay mode (two nucleon absorption, $\bar{K}NN \rightarrow YN$). In all calculations plotted in Fig. \ref{Summary_Latest-Kpp}, such a decay mode is not taken into account, while the mesonic decay mode ($\bar{K}N \rightarrow \pi Y$) is included. Indeed, it is shown in Ref. \cite{2Nabs:Bayar-Oset} that the non-mesonic decay mode gives sizable contribution to the decay width. Recently, we are considering another possible reason for this discrepancy \cite{Semi-rela:Shinmura}: The ``$K^-pp$" is a nuclear system in which an antikaon is bound. However, pion appears in its decay as mentioned above. Most of current studies have been performed in the non-relativistic kinematics. The pion is treated non-relativistically, although it should be treated relativistically because its mass is small. Therefore, the kinetic energy of the pion can easily increase, and accordingly the phase volume for the mesonic decay involving the pion is suppressed. As a result, the decay width might be estimated small in theoretical calculations. We are investigating the relativistic effect to the decay width.

\section{SUMMARY AND FUTURE PROSPECTS}

Kaonic nuclei (nuclear system with antikaons) have been a long-standing issue in strange nuclear physics and hadron physics, because they are expected to have exotic natures such as formation of a dense state which are caused by the strong $\bar{K}N$ attraction. To clarify the nature of kaonic nuclei, many theorists and experimentalists have investigated the most essential system among kaonic nuclei, so-called $K^-pp$, which is a three-body system that would be composed of $K^-$ meson and two protons in a naive consideration. We have also tackled this issue. 

According to most of earlier theoretical studies, the $K^-pp$ is considered to be a resonant state of a coupled-channel system of $\bar{K}NN$, $\pi\Sigma N$ and $\pi\Lambda N$ components. Hence, for the perfect treatment of the $K^-pp$ in a theoretical viewpoint, we have developed a fully coupled-channel complex scaling method (Full ccCSM), respecting both aspects of resonance and coupled channels which are considered important factors in the study of the $K^-pp$. Similarly to earlier studies, we consider the ``$K^-pp$'' system, which represents symbolically the $\bar{K}NN$-$\pi\Sigma N$-$\pi\Lambda N$ coupled-channel system with quantum numbers $J^\pi=0^-$ and $T=1/2$. In the current study, we wish to investigate the $K^-pp$ as theoretically as possible. Therefore, instead of a phenomenological potential employed in our previous study \cite{Full-ccCSM_AY:Dote}, we have used a chiral SU(3)-based $\bar{K}N$(-$\pi Y$) potential, which is based on the chiral theory that is an effective theory of quantum chromodynamics (the fundamental theory for hadron physics). The chiral potential is an energy-dependent potential because of the chiral dynamics. To handle this energy dependence of the potential adequately, we have carried out a self-consistent calculation with an improved prescription for the current coupled-channel case, which was originally proposed for the single-channel case \cite{Kpp:DHW}. On the sharing of the meson binding energy by two meson-baryon bonds in the ``$K^-pp$'', two extreme ansatzes, field picture and particle picture, have been examined in the current study as well as the original study. 

We have successfully carried out self-consistent calculations with Full ccCSM using a chiral SU(3)-based $\bar{K}N$(-$\pi Y$) potential. In other words, self-consistent solutions have been found in all cases that we have examined. When our $\bar{K}N$(-$\pi Y$) potential is constrained with the latest $\bar{K}N$ scattering length which is obtained by the analysis \cite{ChU:IHW} of the precise measurement of the $1s$-level shift energy of kaonic hydrogen atom (SIDDHARTA experiment \cite{Exp:SIDDHARTA}), our results are summarized as follows:

\begin{itemize}
\item The ``$K^-pp$'' is obtained as a shallowly bound state with narrow width if the field picture is employed:  $(B_{K^-pp}, \; \Gamma_{\pi YN}/2) = (14-28, \; 8-15) \, {\rm MeV}$. If the particle picture is employed, it is obtained as a kind of deeply bound state with slightly broad width, compared with the field-picture case: $(B_{K^-pp}, \; \Gamma_{\pi YN}/2) = (21-50, \; 13-19) \, {\rm MeV}$. ($B_{K^-pp}$ and $\Gamma_{\pi YN}$ mean the binding energy of the ``$K^-pp$'' and its mesonic decay width, respectively.)

\item  Since the size of the system is related to its binding energy, the ``$K^-pp$'' obtained with the particle picture is compacter than that obtained with the field picture: as a typical case, the $NN$ mean distance is evaluated as 1.8 fm in the particle picture, whereas it is evaluated as 2.1 fm in the field picture.
\end{itemize}

\noindent 
As described above, the results with our current calculation depend on the ansatz for the self-consistency: the binding energy of the ``$K^-pp$'' is obtained to be smaller with the field picture than with the particle picture. There is about 10-MeV difference in the binding energy between two ansatzes. It is noticed that the result obtained with the field picture is quite similar to the results of earlier studies with chiral-type potentials which suggest that the ``$K^-pp$'' is shallowly bound with the binding energy of $\sim$20 MeV and the $NN$ mean distance of 2.2 fm \cite{Kpp:DHW, Kpp:BGL}. 

Based on the current results of our study, we have made a consideration on the density of kaonic nuclei as follows. It is well known that the mean distance between two nucleons in the normal nuclear matter is 2.2 fm. Compared with this ``normal distance", the above-mentioned $NN$ mean distance indicates that the ``$K^-pp$'' might have the density corresponding to the normal nuclear density $\rho_0$ ($=0.16$ fm$^{-3}$) in case of the field picture, though it might have slightly higher density than the $\rho_0$ in case of the particle picture.  By the way, when a phenomenological $\bar{K}N$ potential \cite{AY_2002} is employed, the $NN$ mean distance of the ``$K^-pp$'' is obtained as 1.8 fm, which corresponds to the density of $1.6\rho_0$. Therefore, if the $\bar{K}N$ potential is sufficiently attractive as a phenomenological potential, kaonic nuclei could be dense matter. However, it is rather probable that kaonic nuclei are normal matter, if the $\bar{K}N$ potential is weakly attractive as a chiral-type potential.  

We have made a comparison of the current results with experimental results searching for the $K^-pp$ quasi- bound state. By a naive comparison, it is quite obvious that our results can not explain the results of earlier experiments such as FINUDA \cite{Kpp:exp_FINUDA}, DISTO \cite{Kpp:exp_DISTO}, and J-PARC E27 \cite{Kpp-ex:JPARC-E27}, which reported signals around 100 MeV below the $\bar{K}NN$ threshold. As for the latest result of J-PARC E15 (second run, J-PARC E15$^{\rm 2nd}$), the reported binding energy, $\sim 50$ MeV \cite{Kpp-ex:JPARC-E15-2nd_fin}, could be explained by the results calculated with phenomenological potential, and somehow by our result obtained with a chiral-type potential employing the particle picture.  However, there is a large discrepancy on the decay width between theoretical results and experimental result. This could be due to the non-mesonic decay mode, which is not taken into account in most of theoretical studies. 

As for future prospects of our study, there still remain several tasks to clarify the nature of kaonic nuclei: 1) At first, we will try to understand the experimental result of J-PARC E15$^{\rm 2nd}$ in more detail, because their data are of high statistics and hence reliable compared with other past experiments. As mentioned in the section of discussion, it is not so easy to compare theoretical results with experimental results, especially in case of resonant states involving broad decay width. For the comparison of theoretical and experimental results, the best is the direct comparison of reaction spectrum (cross section) between them. To do so, the reaction spectrum needs to be calculated in the theoretical side. Fortunately, in the complex scaling method it is unique to construct the Green's function of the system \cite{CSM:Myo2}. In addition, there is a famous method to calculate reaction spectrum with Green's functions, which was proposed by Prof. Morimatsu and Prof. Yazaki \cite{Morimatsu-Yazaki}. With the combination of these two methods, the reaction spectrum will calculated precisely in the theoretical side, and then it can be directly compared to the J-PARC E15$^{\rm 2nd}$ result. 2) As for the decay width, also as mentioned in the discussion, the non-mesonic decay mode, which might give sufficient contribution to the result \cite{2Nabs:Bayar-Oset}, is missed in most of theoretical studies. We will incorporate it into our calculation. In addition, we are going to consider the relativistic effect to the decay width, since it might give some influence to the decay through the phase volume of the pion appeared in the decay process \cite{Semi-rela:Shinmura}. 3) To make our study more accurate, we will examine a sophisticated ansatz for the self-consistency on the meson-baryon energy \cite{Kpp:BGL}, and employ the latest version of chiral SU(3)-based $\bar{K}N$(-$\pi Y$) potential \cite{ChU:MHW}. With these further trials, we will give a definite conclusion to the longstanding issue on the $K^-pp$. Furthermore, by investigating kaonic nuclear systems beyond the $K^-pp$, four-body systems such as $K^-ppn$, $K^-ppp$ and $K^-K^-pp$ (double kaonic nucleus), we would acquire more unified understanding on the nature of kaonic nuclei.

\section{ACKNOWLEDGMENTS}
The corresponding author (A. D.) thanks to the organizers of HYP2018 for the invitation and many supports to attend the conference. In addition, he thanks Prof. T.~Harada, Prof. S.~Shinmura and Prof. Y.~Akaishi for fruitful discussions and Prof. H.~Horiuchi and Prof. H.~Toki for their strong encouragement. This work is supported partially by JSPS KAKENHI Grant Numbers 15K05091, 18K03660 and 26400281. 



\end{document}